\documentclass[fleqn,usenatbib,useAMS]{mnras}

\usepackage{mathtools}

\usepackage{graphicx}	
\usepackage{amsmath}	
\usepackage{amssymb}	
\usepackage{multicol}        
\usepackage{bm}		
\usepackage{pdflscape}	

\newcommand{\rij}{r_{ij}}
\newcommand{\tij}{\theta_{ij}}
\newcommand{\sij}{\sigma_{ij}}

\newcommand{\tilsm}{\tilde{\sigma}}

\newcommand{\nstop}{n_\mathrm{stop}}

\usepackage[T1]{fontenc}
\usepackage{ae,aecompl}

\usepackage{newtxtext,newtxmath}

\title[Geometric cross-sections of fractal aggregates]{Analytic expressions for geometric cross-sections of fractal dust aggregates}

\author[R. Tazaki]{Ryo Tazaki$^{1,2}$
\thanks{E-mail: \href{mailto:r.tazaki@uva.nl}{r.tazaki@uva.nl}}\\
$^{1}$Anton Pannekoek Institute for Astronomy, University of Amsterdam, Science Park 904, 1098XH Amsterdam, The Netherlands\\
$^{2}$Astronomical Institute, Graduate School of Science Tohoku University, 6-3 Aramaki, Aoba-ku, Sendai 980-8578, Japan
}

\pubyear{2021}

\begin{document}
\label{firstpage}
\pagerange{\pageref{firstpage}--\pageref{lastpage}}
\maketitle

%
\begin{abstract}
In protoplanetary discs and planetary atmospheres, dust grains coagulate to form fractal dust aggregates. The geometric cross-section of these aggregates is a crucial parameter characterizing aerodynamical friction, collision rates, and opacities.
However, numerical measurements of the cross-section are often time-consuming as aggregates exhibit complex shapes.
In this study, we derive a novel analytic expression for geometric cross-sections of fractal aggregates. If an aggregate consists of $N$ monomers of radius $R_0$, its geometric cross-section $G$ is expressed as
\begin{eqnarray}
\frac{G}{N\pi R_0^2}=\frac{A}{1+(N-1)\tilsm}, \nonumber
\end{eqnarray}
where $\tilsm$ is an overlapping efficiency, and $A$ is a numerical factor connecting the analytic expression to the small non-fractal cluster limit. The overlapping efficiency depends on the fractal dimension, fractal prefactor, and $N$ of the aggregate, and its analytic expression is derived as well. The analytic expressions successfully reproduce numerically measured cross-sections of aggregates. We also find that our expressions are compatible with the mean-field light scattering theory of aggregates in the geometrical optics limit. The analytic expressions greatly simplify an otherwise tedious calculation and will be useful in model calculations of fractal grain growth in protoplanetary discs and planetary atmospheres.
\end{abstract}





\section{Introduction}\label{sec:intro}

Fractal aggregation is a ubiquitous phenomenon occurring in dust coagulation in nature \citep{Forrest79, Meakin88, Meakin91, Dominik97, Wurm98, Kempf99, Blum00, Krause04, P06, S08, Wada08, S12}. It is a primary mode of grain growth in star- and planet-forming regions \citep{O93, W93, Ormel07, Ormel09, O12, K13, Krijt15, Krijt16}, which is supported by findings of such aggregates in comet 67P/Churyumov-Gerasimenko \citep{Bentley16, Mannel16}. Cloud and haze particles in planetary atmospheres also form fractal aggregates, such as in the Early Earth \citep{Wolf10}, Titan \citep{Cabane93, Rannou95, Rannou97, Tomasko08}, Pluto and Triton \citep{Lavvas20, Ohno20b}, and exoplanets \citep{M13, Ohno20a}. 

Influences of fractal grain growth are manifold, such as planetesimal formation \citep{O12, K13}, the radiation environment of the Early Earth \citep{Wolf10}, and the observational appearance of protoplanetary discs \citep{T19a,T19b} and exoplanetary atmospheres \citep{Ohno20a}. Thus, it is crucial to learn how fractal dust aggregates form and grow in these environments. The geometric cross-section, or the projected area, of a fractal aggregate is a key to elucidate its dynamical evolution, as it determines aerodynamical friction and collision rates.

\begin{figure}
\begin{center}
\includegraphics[width=\linewidth,keepaspectratio]{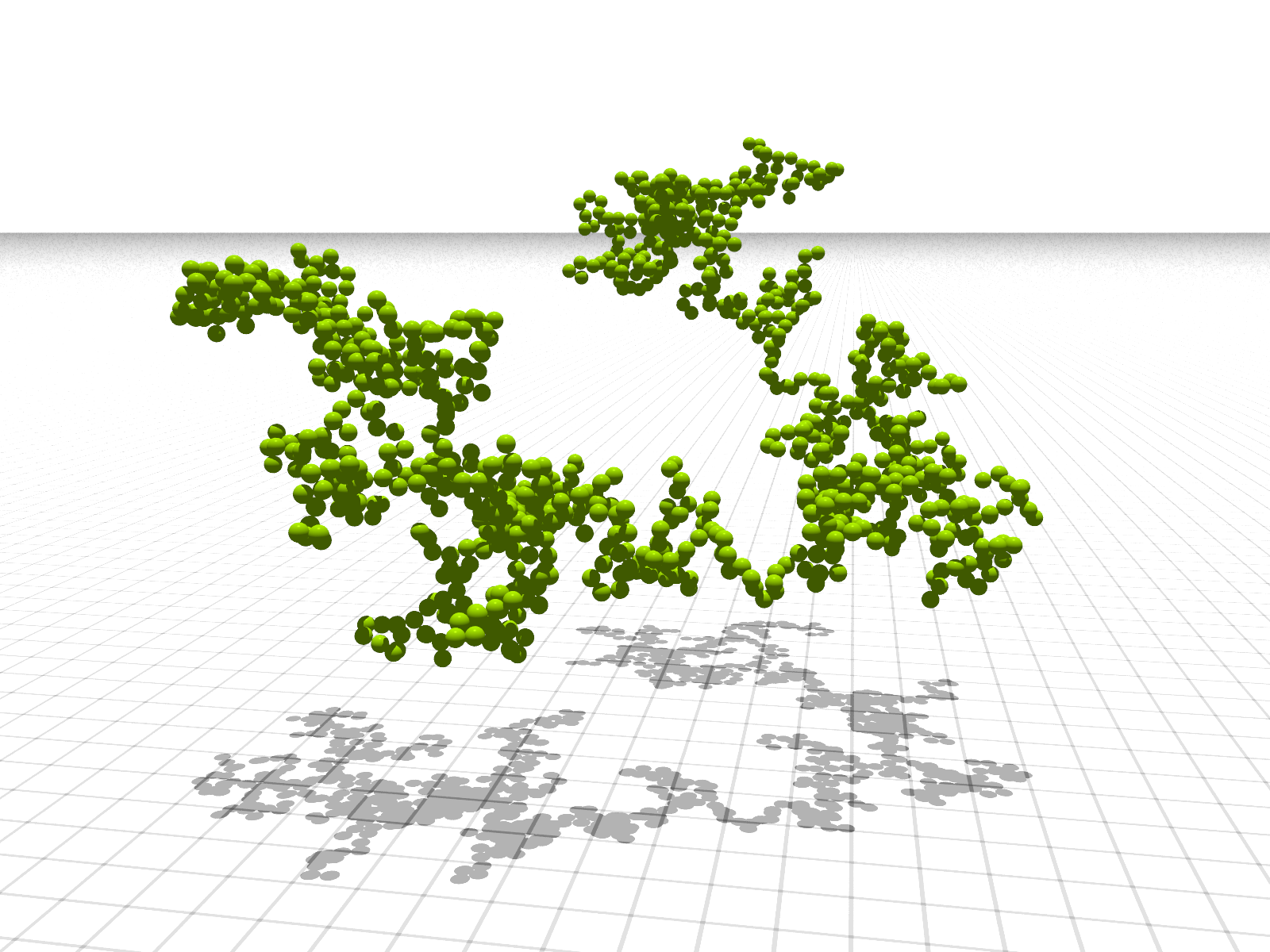}
\caption{A fractal dust aggregate formed via BCCA. The shadow on the ground depicts the geometric cross-section of the aggregate.}
\label{fig:bcca_shadow}
\end{center}
\end{figure}

Fractal aggregates typically exhibit complex shapes (Figure \ref{fig:bcca_shadow}). Thus, numerical simulations have been invoked to evaluate their cross-sections (e.g., \citealp{Meakin88, Mukai92, O93, Kempf99, Minato06}; Okuzumi, Tanaka, \& Sakagami 2009, hereafter \citealp{O09}; \citealp{P09, S12}). These are, however, time-consuming for large aggregates, as a sufficient number of orientation and ensemble averages must be taken into account to suppress statistical fluctuations. Although several empirical formulas are present \citep{Meakin88, O93, Minato06, P09, O09}, an analytic expression that is valid for the entire range of fractal dimensions has not been derived to date. 

Meanwhile, an analytic light scattering theory of fractal aggregates, based on a statistical distribution model of constituent particles (henceforth called monomers), has proven to be successful \citep{Berry86, B97, Rannou97, T16, T18}. This success motivated us to develop an analytic model of geometric cross-sections of fractal aggregates based on a similar statistical approach. To our knowledge, this is the first attempt at this type of approach. 

This study presents a novel analytic model of geometric cross-sections of fractal aggregates by applying a statistical distribution model of monomers. We demonstrate that our analytic model successfully reproduces numerically measured cross-sections reported in earlier studies \citep[e.g.,][]{Meakin88, Minato06, O09}. Moreover, our model is shown to be compatible with the mean-field light scattering theory \citep{B97}. 

This paper is organized as follows.
In Section \ref{sec:scaling}, we summarize scaling properties of geometric cross-sections of fractal aggregates based on earlier studies. In Section \ref{sec:ana}, we derive an analytic expression for geometric cross-sections and test its validity by comparison with numerical results reported in previous studies. A relationship between the analytic expression and the mean-field light scattering theory is discussed in Section \ref{sec:mft}. 
In Section \ref{sec:disc}, we further extend the analytic expression to treat the cross-sections of inhomogeneous aggregates and test its validity. Also, we discuss the applicability of empirical formulas that have been commonly used in previous studies. 
A summary is given in Section \ref{sec:con}.

\section{Scaling properties of geometric cross-sections} \label{sec:scaling}
\subsection{Essential parameters describing fractal aggregates}

The structure of a fractal aggregate is characterized by three parameters: 
the fractal dimension $D$, fractal prefactor $k_0$, and the number of monomers $N$. The radius of gyration is commonly used to describe an aggregate radius and obeys a well-known fractal law:
\begin{equation}
R_g=R_0\left(\frac{N}{k_0}\right)^{1/D}, \label{eq:Rg}
\end{equation}
where $R_0$ is the monomer radius. The characteristic radius of the aggregate, $R_c=\sqrt{5/3}R_g$, has likewise been employed \citep{Kozasa92, Mukai92}, instead of the radius of gyration. In Table \ref{tab:rad}, we summarize some definitions of aggregate radii used in this paper.

The values of $D$ and $k_0$ depend on aggregation processes. 
A primary stage of aggregation often occurs by a process called cluster-cluster aggregation (CCA). CCA results in forming aggregates with fluffy structure (Figure \ref{fig:bcca_shadow}). Numerical simulations suggest that a CCA cluster typically has $D=1.7-2.1$ \citep[e.g.,][]{Meakin84a, Meakin84b, B95, Sorensen97, Kempf99}. A fractal dimension of a non-ballistic CCA cluster can be even lower, namely as low as $D\sim1.1-1.4$, owing to the effect of rotation of aggregates during collisions \citep{Blum00, Krause04, P06}. Ballistic particle-cluster aggregation (BPCA) offers the opposite limiting case to CCA, as it results in forming nearly homogeneous aggregates with $D\sim3.0$ with approximately 85 per cent porosity \citep{Kozasa92}. 

In this study, we consider the following three well-known fractal aggregation models:  ballistic-CCA (BCCA), BPCA, and linear chain, where each model exhibits ($D=1.9,~k_0=1.04$), ($D=3.0,~k_0=0.3$), ($D=1.0,~k_0=\sqrt{3}$), respectively \citep[e.g.,][]{T16, Sorensen97}.
The linear chain cluster may be regarded as an analogue of aggregates formed via non-ballistic CCA. 
The fractal dimensions and fractal prefactors of these aggregates approximately follow the linear anti-correlation:
\begin{equation}
k_0(D) \approx 0.716(1-D)+\sqrt{3}. \label{eq:k0app}
\end{equation}
Equation (\ref{eq:k0app}) is linear interpolation between $D=1.0$ and $D=3.0$ and reproduces the measured $k_0$ value of the BCCA model within the error of $5$ per cent. 

\begin{table}
  \caption{A summary of symbols and definitions for various dust radii}
  \label{tab:rad}
  \centering
  \begin{tabular}{lll}
    \hline
    Symbol & Definition & Meaning \\
    \hline \hline
    $R_0$ & - & Monomer radius  \\
    $R_g$ & $R_0(N/k_0)^{1/D}$ & Radius of gyration  \\
     $R_c$ & $\sqrt{5/3}R_g$ & Characteristic radius  \\
     $R_a$ & $\sqrt{G/\pi}$ & Area-equivalent radius  \\
    \hline
  \end{tabular}
\end{table}

\subsection{Scaling properties of geometric cross-sections} \label{sec:sl}
In general, the geometric cross-section depends on orientation of a fractal aggregate (see e.g., Figure \ref{fig:bcca_shadow}). 
The cross-section also depends on the formation history of each aggregate.
For example, BCCA clusters consisting of $N$ monomers of radius $R_0$ exhibit various sizes and shapes. Meanwhile, model calculations of fractal grain growth often require an average cross-section of these aggregates rather than a cross-section measured at a specific orientation and a formation history. Therefore, it is reasonable to take the average of the cross-sections across various orientations for each aggregate and an ensemble of aggregates with different formation histories. Let $G$ denote this averaged geometric cross-section. Hereafter, we only focus on how the average cross-section $G$ is expressed as a function of $R_0,~D,~k_0$, and $N$.

The simplest approach for estimating the cross-section is to employ {\it the characteristic cross-section}: 
\begin{equation}
G=\pi R_c^2. \label{eq:pirc2}
\end{equation}
It is useful to normalize $G$ by $N\pi R_0^2$, where $N\pi R_0^2$ is the sum of the cross-sections of individual monomers, and therefore, in general, we have $G/N\pi R_0^2\le1$. 
By using the normalization, Equation (\ref{eq:pirc2}) scales as 
\begin{equation}
\frac{G}{N\pi R_0^2}\propto \frac{1}{N}\left(\frac{R_g}{R_0}\right)^2\propto N^{\frac{2}{D}-1}. \label{eq:s1}
\end{equation}
Equation (\ref{eq:s1}) demonstrates that the characteristic cross-section is problematic when $D<2$. In this case, $G/N\pi R_0^2$ increases with $N$ and eventually exceeds unity, which is clearly unphysical.

\citet{Minato06} measured geometric cross-sections of BCCA and BPCA clusters and derived fitting formulas. For $N<16$, the two types of clusters exhibit approximately the same cross-section
\begin{equation}
\frac{G}{N\pi R_0^2} =12.5N^{-0.315}\exp\left(-2.53/N^{0.0920}\right), \label{eq:minatosmall}
\end{equation}
and for $16\le N\le10^4$,
\begin{equation}
\frac{G}{N\pi R_0^2}=
\begin{dcases}
4.27N^{-0.315}\exp\left(-1.74/N^{0.243}\right) & (\mathrm{BPCA}),\\
0.352+0.566N^{-0.138} & (\mathrm{BCCA}).
\end{dcases} \label{eq:minato}
\end{equation}

For BPCA, Equation (\ref{eq:minato}) yields $G/N\pi R_0^2\propto N^{-0.315}$ for large $N$. A similar exponent has been derived in previous studies: \citet{Mukai92} derived the exponent of $\le-0.302$, and \citet{O93} found $-1/3$. These exponents seemingly agree with that predicted by the characteristic cross-section, $-1/3$ for $D=3$ in Equation (\ref{eq:s1}). 

For BCCA, Equation (\ref{eq:minato}) yields $G/N\pi R_0^2=$ const. for large $N$. Previous studies have also found the same property \citep{Meakin88, Mukai92}. In addition, this scaling property is valid for even lower $D$ values: $D\sim1.3$ \citep{Blum00} and $D=1.0$ (see Appendix \ref{sec:chain}). 
Perhaps, weak logarithmic dependence may be present for $D\sim2$ \citep[e.g.,][]{Meakin88,O93}. These dependencies are incompatible with the characteristic cross-section.

Therefore, according to the numerical measurements, it is expected that the geometric cross-sections of fractal dust aggregates with {\it sufficiently large $N$} scale as
\begin{equation}
\frac{G}{N\pi R_0^2} \propto
\begin{dcases}
\mathrm{const.} & (D<2),\\
\frac{1}{\ln{N}} & (D=2),\\
N^{-1}R_g^2\propto N^{\frac{2}{D}-1} & (2< D\le3).
\end{dcases} \label{eq:scaling}
\end{equation}

At this point, this scaling law is purely empirical. However, our physically motivated analytic expressions, derived in Section \ref{sec:ana}, provide a clear justification for this scaling law.

\section{Analytic expressions for geometric cross-sections} \label{sec:ana}

We derive analytic expressions for geometric cross-sections of fractal aggregates.
\subsection{Geometric cross-section of a distribution of monomers} \label{sec:derive}
\begin{figure}
\begin{center}
\includegraphics[width=\linewidth,keepaspectratio]{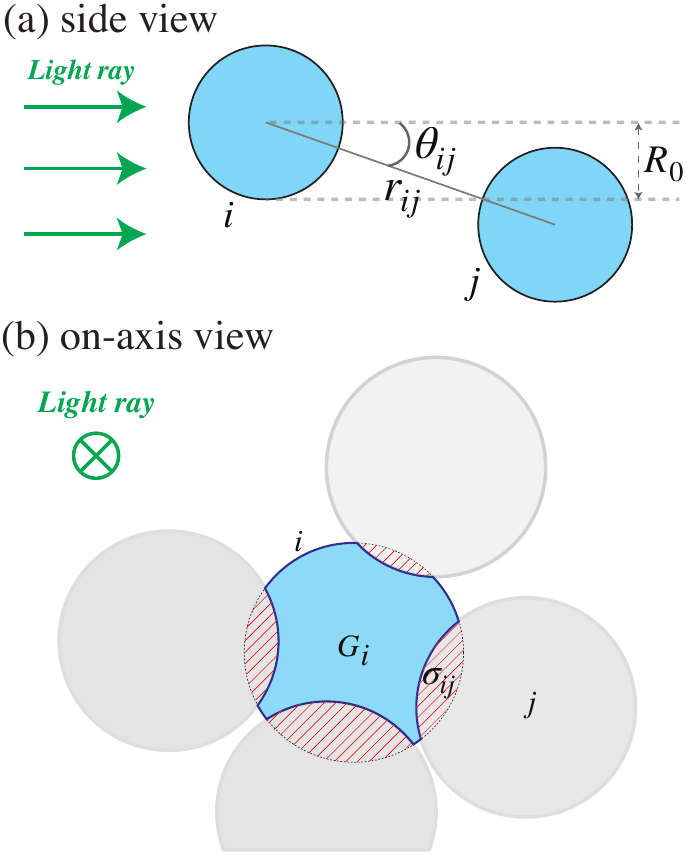}
\caption{Cartoons illustrating overlapping monomers. (a) A two-monomer system, where one ($i$th) casts a shadow on the other ($j$th) from a side view of the light ray. (b) A part of an aggregate from an on-axis view of the light ray. In this case, the $i$-th monomer overlaps with four monomers. $G_i$ represents the unshaded area of the $i$-th monomer. The shaded region indicated by $\sij$ represents the area of the overlapping region of the spheres $i$ and $j$.}
\label{fig:cartoon1}
\end{center}
\end{figure}

We start with the simplest case, namely, the case of two spheres (Figure \ref{fig:cartoon1}a). If the two spheres overlap along a light ray direction, one sphere (the $i$th sphere) casts a shadow onto the other one (the $j$th sphere). 
By defining $\sij$ as the overlapping area of the two spheres in projection onto the plane perpendicular to the light direction, it is thus given by
\begin{equation}
\sij=2R_0^2 \left[\arcsin\left(\sqrt{1-\rho^2u^2}\right)-\rho u \sqrt{1-\rho^2u^2}\right], \label{eq:sigij}
\end{equation}
where $\rho=\rij/2R_0$, $u=\sin\tij$, and $\rij$ and $\tij$ denote the distance and angle between the two spheres, respectively (see Figure \ref{fig:cartoon1}). For the two spheres to overlap, $0\le \rho u\le1$ is required.

Next, we derive an analytic expression for the geometric cross-section of an aggregate of $N$ monomers. Although we will derive this expression in a more general way later (Section \ref{sec:mft}), below, we first motivate the expression by deriving it for a simple case without multiple overlaps of monomers along a light ray and then proceed to demonstrate its general applicability.

If $\sum_{j\ne i}\sigma_{ij}\ll \pi R_0^2$ and monomer pair distribution can be regarded as isotropic, multiple overlaps would be sub-dominant, and therefore, every overlapping region on a monomer is likely to be isolated each other (Figure \ref{fig:cartoon1}b). In this case, the unshaded area of the $i$-th monomer is 
\begin{equation}
G_i=\pi R_0^2 - \sum_{j\ne i}\sigma_{ij}. \label{eq:gi}
\end{equation}
By using $G_i$, the geometric cross-section of an aggregate may be approximated by 
\begin{equation}
G\simeq \sum_{i=1}^{N}G_i, \label{eq:gsum}
\end{equation}
where we have neglected a contribution of the overlapping regions to the cross-section. 

To evaluate $\sum_{j\ne i}\sigma_{ij}$ in Equation (\ref{eq:gi}), we consider the average overlapping area across all pairs ${\sigma}=\left<\sigma_{ij}\right>$ and assume that the overlap occurring in each pair $ij$ is equal to the average value: $\sigma_{ij}={\sigma}$. 
We further introduce a normalized version of the average overlapping area ${\tilde \sigma}={\sigma}/\pi R_0^2$, which is referred to as {\it the overlapping efficiency} in this paper. 
With these assumptions and definitions, Equations (\ref{eq:gi} and \ref{eq:gsum}) become
\begin{equation}
\frac{G}{N\pi R_0^2} \simeq 1 - (N-1)\tilsm. \label{eq:gweak}
\end{equation}
Equation (\ref{eq:gweak}) is valid only when $(N-1)\tilsm\ll1$, and hence, it may be rewritten as
\begin{equation}
\frac{G}{N\pi R_0^2} \simeq \frac{1}{1+(N-1){\tilsm}}.\label{eq:sigmean}
\end{equation}

Equation (\ref{eq:sigmean}) might be applicable even when multiple overlaps are dominant, i.e., $(N-1)\tilsm>1$. For example, in the opposite extreme case for fully overlapping monomers in a linear chain cluster ($\tilsm=1$), Equation (\ref{eq:sigmean}) yields a correct geometric cross-section $G=\pi R_0^2$ regardless of a value of $N$. Thus, this equation seems to capture the cross-section in both limiting cases.

Equation (\ref{eq:sigmean}) can also be justified in terms of a light scattering theory of aggregates. In Section \ref{sec:mft}, we obtain an identical expression to Equation (\ref{eq:sigmean}) by considering the geometrical optics limit of the mean-field light scattering theory \citep{Berry86, B97}. Thus, we use Equation (\ref{eq:sigmean}) as a basic equation to compute geometric cross-sections of aggregates.

To find $\tilsm$, we employ a statistical distribution model of monomers in fractal aggregates, namely, the two-point correlation function \citep[e.g.,][]{Meakin91, B95, T16, T18}.
If the distribution of monomer pairs is isotropic, the correlation function depends only on the relative distance of a pair $r$. In this case, the form of the correlation function for fractal aggregates is \citep[e.g.,][]{B97, T18}
\begin{eqnarray}
g(r)dr&=&\frac{1}{4\pi R_g^3}\left(\frac{r}{R_g}\right)^{D-3}
f_c\left(\frac{r}{R_g}\right)dr, \label{eq:gumodel}\\
f_c\left(\frac{r}{R_g}\right)&=&\frac{D}{2}\exp\left[-\frac{1}{2}\left(\frac{r}{R_g}\right)^D\right], \label{eq:comodel1}
\label{eq:comodel}
\end{eqnarray}
where $g(r)dr$ yields the probability of finding a pair of monomers separated by a distance between $r$ and $r+dr$. Equation (\ref{eq:gumodel}) is normalized such that $\int_0^\infty 4\pi r^2 g(r)dr=1$. 

Although Equation (\ref{eq:gumodel}) is a general expression describing fractal aggregates \citep[e.g.,][]{Meakin91}, the form of the cut-off function (Equation \ref{eq:comodel}) is non-trivial. This study adopts a model with the cut-off power of $r^D$ as a fiducial case, as suggested by \citet{B95}. \citet{T18} confirmed that this cut-off model is an appropriate choice for BCCA and BPCA. Another possible choice is a model with the cut-off power of $r^2$ \citep[e.g.,][]{T16}, which provides results that are only slightly less accurate than those of the model of $r^D$.
However, a model with the cut-off power of $r$ \citep{Berry86} is significantly inaccurate compared to the other models. The detailed comparison between these three cut-off models is presented in Appendix \ref{sec:cutoff}.

Employing the two-point correlation function, the average overlapping area is given by
\begin{equation}
\langle \sij \rangle = 2\pi \int_{2R_0}^\infty d\rij \rij^2 g(\rij)  \int_0^{\arcsin(1/\rho)}  d\tij \sij \sin\tij. 
\end{equation}
By replacing the integration variables from ($\rij$, $\tij$) to ($\rho$, $u$), the overlapping efficiency becomes
\begin{eqnarray}
\tilsm&=&32 \int_1^\infty d\rho \rho^2 R_0^3g(\rho)\int_0^{1/\rho} du \frac{u}{\sqrt{1-u^2}}\nonumber\\ 
&&\left[\arcsin\left(\sqrt{1-\rho^2u^2}\right)-\rho u\sqrt{1-\rho^2u^2}\right].  \label{eq:tilsm_gen}
\end{eqnarray}

A further analytical reduction of Equation (\ref{eq:tilsm_gen}) is possible by assuming $\rho\gg1$, i.e., the distance between two monomers is larger than the monomer diameter.
In this case, the integration with respect to $u$ is approximated as
\begin{eqnarray}
&&\int_0^{1/\rho} du\frac{u}{\sqrt{1-u^2}}\left[\arcsin\left(\sqrt{1-\rho^2u^2}\right)-\rho u\sqrt{1-\rho^2u^2}\right]\nonumber\\
&\simeq& \int_0^{1/\rho}du~u\left[\arcsin\left(\sqrt{1-\rho^2u^2}\right)-\rho u \sqrt{1-\rho^2u^2}\right], \label{eq:angapp}\\
&=&\frac{\pi}{16\rho^2}, 
\end{eqnarray}
where we used $u\ll1$, since $\rho u\le1$. In general, this approximation tends to be inaccurate for lower $D$. 
However, even for such cases, this approximation only slightly affects accuracy of $G$, since these aggregates tend to have lower overlapping efficiency (i.e., $N\tilsm\lesssim1$), and then the error of $\tilsm$ weakly affects a resultant value of $G$ (see Section \ref{sec:test}).

Hence, Equation (\ref{eq:tilsm_gen}) is reduced to
\begin{equation}
\tilsm\simeq2\pi \int_1^\infty d\rho R_0^3g(\rho). \label{eq:tsmint}
\end{equation}
By using Equations (\ref{eq:gumodel}) and (\ref{eq:comodel}), we obtain
\begin{equation}
\tilsm=\frac{\eta^{2/D}}{16}\int_{\eta}^\infty x^{-\frac{2}{D}}e^{-x}dx,
 \label{eq:tilsig_int}
\end{equation}
where 
\begin{equation}
\eta=2^{D-1}\left(\frac{R_0}{R_g}\right)^D=2^{D-1}\frac{k_0}{N}. \label{eq:eta}
\end{equation}

When $D\le2$, numerical integration is necessary to evaluate $\tilsm$. However, when $2<D\le3$, we have a more convenient expression:
\begin{equation}
\int_{\eta}^\infty x^{-\frac{2}{D}}e^{-x}dx=\Gamma\left(\frac{D-2}{D}\right)Q\left(\frac{D-2}{D},\eta\right),
\end{equation}
where $\Gamma$ is the Gamma function, and $Q$ is the incomplete Gamma function defined by 
\begin{equation}
Q(a,x)\equiv\frac{1}{\Gamma(a)}\int_x^\infty t^{a-1}e^{-t} dt,
\end{equation}
where $a>0$ \citep{AS72}. $Q(a,x)$ can be efficiently computed without numerical integration \citep{Press92}. This is the main practical reason for using the incomplete Gamma function rather than performing the numerical integration.

Consequently, the analytic expressions for the overlapping efficiency are expressed as
\begin{equation}
\tilsm=\frac{\eta^{2/D}}{16}
\begin{dcases}
\Gamma\left(\frac{D-2}{D}\right)Q\left(\frac{D-2}{D},\eta\right) & (2<D\le3),\\
\int_{\eta}^\infty dx x^{-\frac{2}{D}}e^{-x} & (D\le2).
\end{dcases} \label{eq:tilsm_final}
\end{equation}

\subsection{Connection to small non-fractal cluster limit} \label{sec:nflim}

The fractal scaling law (Equation \ref{eq:Rg}) breaks down when $N$ is small. In such small clusters, geometric cross-sections would be almost independent of their clustering processes, and therefore the formal fractal dimension.
In fact, \citet{Minato06} employed the same fitting formula for the cross-sections of BCCA and BPCA clusters when $N<16$ \citep[see also][]{Mukai92, O93}. In contrast, our analytic expression above was derived on the premise that the fractal scaling law is valid even for a small number of monomers, and therefore, it depends on $D$ even for the small cluster limit. 

To reconcile this problem, we adopt a piecewise approach.
For $N<N_\mathrm{th}$, where $N_\mathrm{th}$ is a small number (e.g., below approximately ten), it is reasonable to assume that geometric cross-sections are nearly independent of aggregation processes, and hence, we may use Equation (\ref{eq:minatosmall}). Although \citet{Minato06} practically adopted $N_\mathrm{th}=16$ in their fitting formulas, a slight difference between measured cross-sections of BCCA and BPCA clusters can be seen at even smaller aggregates, $N\sim8$ \citep{Mukai92, O93, Minato06}. 
Thus, in this study, we adopt $N_\mathrm{th}=8$ for BCCA and BPCA.
If we consider further lower $D$ values, $N_\mathrm{th}$ can be even smaller. For example, The cross-sections of linear chain clusters ($D=1$) start to deviate from those predicted by Equation (\ref{eq:minatosmall}) at $N\sim3$. Taking this into consideration, we empirically adopt
\begin{equation}
N_\mathrm{th} = \mathrm{min}\left(11D-8.5, 8.0\right). \label{eq:nth}
\end{equation}
Equation (\ref{eq:nth}) yields $N_\mathrm{th}=8.0$ and $N_\mathrm{th}=2.5$ for $D\ge1.5$ and $D=1.0$, respectively. $N_\mathrm{th}$ for $1.0<D<1.5$ is determined by linear interpolation.

\subsection{Summary of our analytic expression} \label{sec:final}
Our analytic expressions for the geometric cross-section of a fractal aggregate are summarized to be
\begin{equation}
\frac{G}{N\pi R_0^2} =
\begin{dcases}
12.5N^{-0.315}\exp\left(-2.53/N^{0.0920}\right) & (N<N_\mathrm{th}),\\
\frac{A}{1+(N-1)\tilsm} & (N_\mathrm{th} \le N),
\end{dcases} \label{eq:final}
\end{equation}
where $A$ is the numerical factor that connects two regimes continuously at $N=N_\mathrm{th}$.
$\tilsm$ and $N_\mathrm{th}$ are given by Equations (\ref{eq:tilsm_final}) with (\ref{eq:eta}) and Equation (\ref{eq:nth}), respectively.

Our formulation does not involve any computationally challenging tasks; thus, the computational time is almost negligible regardless of values of $N$ and $D$. The source codes are publicly available at the author's GitHub repository \footnote{\url{https://github.com/rtazaki1205/geofractal}}.

\subsection{Tests of our analytic expression} \label{sec:test}
\begin{figure}
\begin{center}
\includegraphics[width=\linewidth]{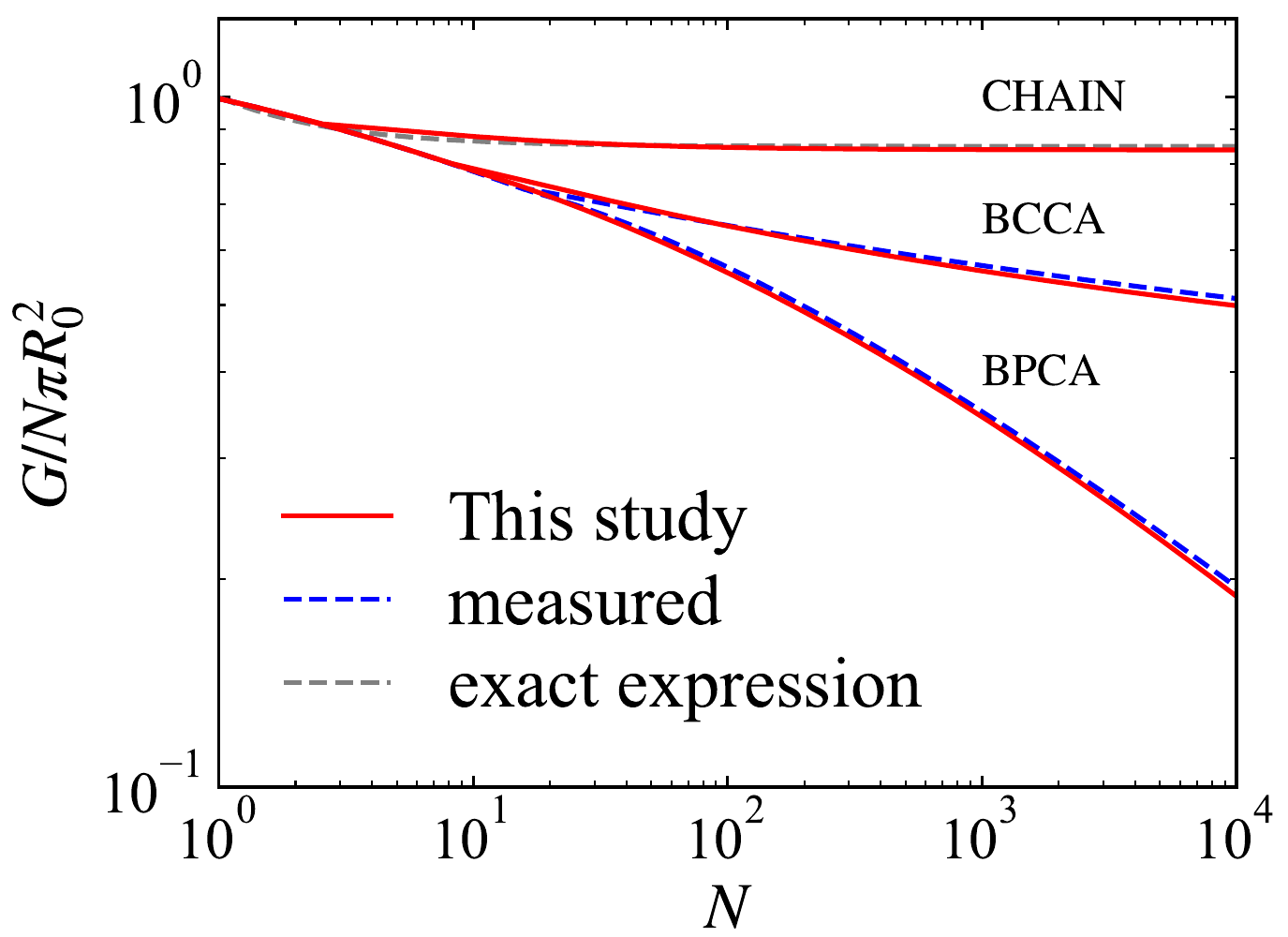}
\caption{Red solid lines represent geometric cross-sections of the linear chain, BCCA, and BPCA clusters obtained by our analytic expression (Equation \ref{eq:final}). Blue and grey dashed lines represent the results obtained by the fitting formulas for BCCA and BPCA clusters \citep{Minato06} and the exact expression for the linear chain cluster (Equation \ref{eq:chainexact}), respectively.}
\label{fig:comp}
\end{center}
\end{figure}

To test the validity of our analytic expression (Equation \ref{eq:final}), we compare it with the fitting formulas for measured cross-sections of BCCA and BPCA clusters (Equations \ref{eq:minatosmall} and \ref{eq:minato}). We also compare it with the exact expression for the cross-section of a linear chain cluster (see Appendix \ref{sec:chain}), which can be written by
\begin{equation}
\frac{G}{N\pi R_0^2}=\frac{1}{N}+\left(1-\frac{1}{N}\right)\frac{8}{3\pi}. \label{eq:chainexact}
\end{equation}

Figure \ref{fig:comp} compares our analytic expression (Equation \ref{eq:final}) with measured or exact cross-sections for the three types of fractal dust aggregates. 
Our expression successfully reproduces the geometric cross-sections for all three types of aggregates. 
The relative error of our expression is only less than 3 per cent (see also Figure \ref{fig:cutoff}). We also compared our expression with a fitting formula for the cross-sections of BCCA clusters proposed by \citet{Meakin88}. 
As a result, we found our expression agrees with the fitting formula within the error of 4.5 per cent for $N<10^8$, where we used $D=1.95$ and $k_0=1.0$ in the comparison \citep{Meakin84a}. The above results demonstrate the validity of our formulation.

We conducted another test to confirm the validity of the approximation made in Equation (\ref{eq:angapp}). Consequently, we found that this approximation causes a relative error below $\sim1$ per cent for $1\le D\le3$ and $1\le N\le 10^{10}$, where we used Equation (\ref{eq:k0app}) to find $k_0$ for each value of $D$. Therefore, the approximation is applicable for a wide parameter space.

\subsection{Scaling properties of our analytic expression}
\begin{figure}
\begin{center}
\includegraphics[width=\linewidth]{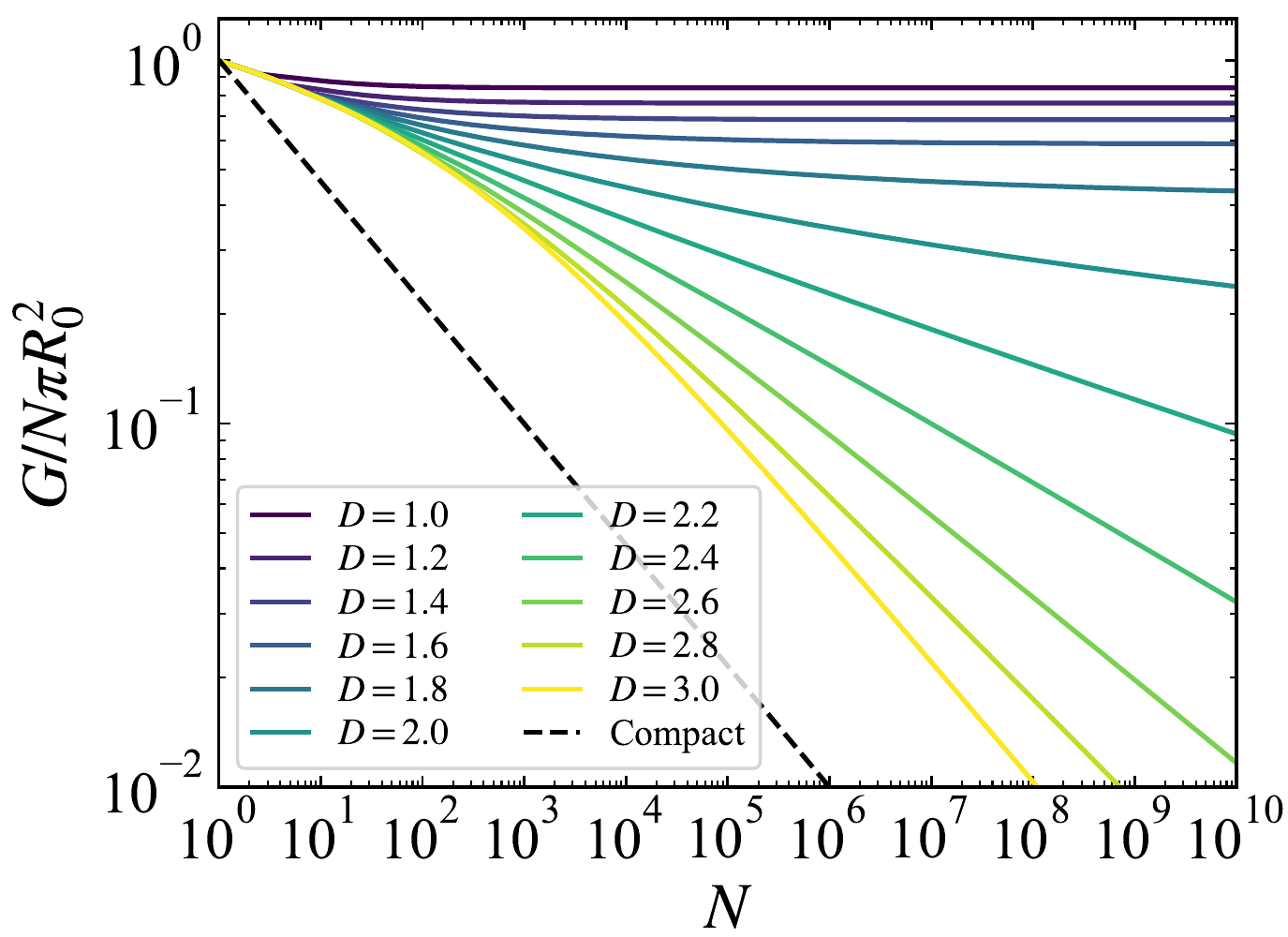}
\caption{Geometric cross-sections of fractal aggregates for various values of $D$ and $N$ obtained by our analytic expression (Equation \ref{eq:final}). The fractal prefactor is determined by Equation (\ref{eq:k0app}). The dashed line represents the cross-sections of non-porous compact spheres.}
\label{fig:df}
\end{center}
\end{figure}

One of the advantages of our analytically derived expression is its applicability to the full range of fractal dimensions. To demonstrate this, Figure \ref{fig:df} shows geometric cross-sections of aggregates obtained by our expression for various values of $D$ and $N$. 
As a general tendency, a higher fractal dimension yields a higher overlapping efficiency, and consequently, a smaller geometric cross-section. For comparison, we plotted the cross-section of non-porous compact spheres, which obey $G/N\pi R_0^2=N^{-1/3}$. Although both BPCA clusters and compact spheres have $D=3.0$, BPCA clusters exhibit significantly larger cross-sections than compact spheres.

Depending on $D$, the slope with respect to $N$ changes at sufficiently large $N$. To assess whether our expression satisfies the scaling law expected in Equation (\ref{eq:scaling}), we investigate the asymptotic behaviour of the analytic expression. The asymptotic behaviour can be classified into three cases (i) $2<D\le3$, (ii) $D=2$, and (iii) $D<2$. 

\subsubsection{Case of $2<D\le3$} \label{sec:asym1}
For sufficiently large aggregates, $Q\simeq1$. Hence, Equation (\ref{eq:tilsm_final}) is reduced to
\begin{equation}
\tilsm\simeq\frac{(R_0/R_g)^2}{2^{2(D+1)/D}}\Gamma\left(\frac{D-2}{D}\right).\label{eq:tilsig_Q1}
\end{equation}
Thus, the geometric cross-section scales as
\begin{equation}
\frac{G}{N\pi R_0^2}\propto\frac{1}{N}\left(\frac{R_g}{R_0}\right)^2\propto N^{\frac{2}{D}-1}.
\end{equation}
This scaling property is in harmony with Equation (\ref{eq:scaling}).

\subsubsection{Case of $D=2$} \label{sec:asym2}
In this case, Equation (\ref{eq:tilsig_int}) gives rise to the exponential integral. To assess its asymptotic behaviour approximately, we omit the cut-off function in the integrand in Equation (\ref{eq:tilsig_int}) and truncate the upper bound of the integral at $\sim1$.
As a result, Equation (\ref{eq:tilsig_int}) becomes
\begin{equation}
\tilsm\propto \left(\frac{R_0}{R_g}\right)^2\ln\left(\frac{R_g}{R_0}\right).
\end{equation}
Therefore, the geometric cross-section scales as
\begin{equation}
\frac{G}{N\pi R_0^2}\propto\frac{1}{\ln{N}}.
\end{equation}
This result is consistent with Equation (\ref{eq:scaling}).

\subsubsection{Case of $D<2$}
Similar to the case of $D=2$, we omit the cut-off function to estimate its asymptotic behaviour. In this case, Equation (\ref{eq:tilsig_int}) becomes
\begin{equation}
\tilsm\propto \left(\frac{R_0}{R_g}\right)^D.
\end{equation}
Therefore, the geometric cross-section scales as
\begin{equation}
\frac{G}{N\pi R_0^2}\propto\frac{1}{N}\left(\frac{R_g}{R_0}\right)^D=\mathrm{const.}
\end{equation}
Therefore, this is also consistent with Equation (\ref{eq:scaling}). 

\section{Short-wavelength limit of mean-field light scattering solution} \label{sec:mft}
In Section \ref{sec:ana}, we derived an analytic expression for geometric cross-sections of aggregates by considering a simple case without multiple overlaps of monomers. 
Another possible approach to derive the expression is to take the short-wavelength limit of the mean-field light scattering theory \citep{B97}. 
The mean-field theory successfully reproduces the extinction cross-sections of BCCA and BPCA clusters \citep{T18}. Since the extinction cross-section approaches twice the geometric cross-section at a sufficiently short wavelength \citep[e.g.,][]{BH83}, it can be used to {\it measure} the geometric cross-section. In this section, we re-derive Equation (\ref{eq:sigmean}) by considering the short-wavelength limit of the mean-field theory.
 
\subsection{Summary of mean-field equations}
Here, we summarize basic equations of the mean-field theory \citep{B97}.
The extinction cross-section of an aggregate is given by
\begin{equation}
C_{\mathrm{ext}}=\frac{2\pi N}{k^2}\sum_{n=1}^{\nstop}(2n+1)\mathrm{Re}\left(\bar{d}_{1,n}^{(1)}+\bar{d}_{1,n}^{(2)}\right) \label{eq:Cextmft},
\end{equation}
where $k$ is the wavenumber, and ($\bar{d}_{1,n}^{(1)}, \bar{d}_{1,n}^{(2)}$) are the mean-field scattering coefficients obtained by solving a set of linear equations \citep{B97} 
\begin{eqnarray}
\bar{d}_{1,n}^{(1)}&=&a_n\left[1-(N-1)\sum_{\nu=1}^{\nstop} \bar{A}_{1,n}^{1,\nu}\bar{d}_{1,\nu}^{(1)}+\bar{B}_{1,n}^{1,\nu}\bar{d}_{1,\nu}^{(2)}\right] \label{eq:basic1},\\
\bar{d}_{1,n}^{(2)}&=&b_n\left[1-(N-1)\sum_{\nu=1}^{\nstop} \bar{B}_{1,n}^{1,\nu}\bar{d}_{1,\nu}^{(1)}+\bar{A}_{1,n}^{1,\nu}\bar{d}_{1,\nu}^{(2)}\right] \label{eq:basic2},
\end{eqnarray}
where  $\bar{A}_{1,n}^{1,\nu}$ are $\bar{B}_{1,n}^{1,\nu}$ defined by
\begin{eqnarray}
\bar{A}_{1,n}^{1,\nu}&=&\frac{2\nu+1}{n(n+1)\nu(\nu+1)}\sum_{p=|n-\nu|}^{n+\nu}[n(n+1)+\nu(\nu+1)\nonumber\\
&&-p(p+1)]a(\nu,n,p)s_p(kR_g), \label{eq:coefA}\\
\bar{B}_{1,n}^{1,\nu}&=&2\frac{2\nu+1}{n(n+1)\nu(\nu+1)}\sum_{p=|n-\nu|}^{n+\nu} b(\nu,n,p)s_p(kR_g),\label{eq:coefB}\nonumber\\\\
a(\nu,n,p)&=&\frac{2p+1}{2}\int_{-1}^{1} P_{\nu}^{1}(x)P_{n}^{1}(x)P_{p}(x)dx \label{eq:anunp}, \\
b(\nu,n,p)&=&\frac{2p+1}{2}\int_{-1}^{1} P_{\nu}^{1}(x)P_{n}^{1}(x)\frac{dP_{p}(x)}{dx}dx \label{eq:bnunp},\\
s_p(kR_g)&=&\frac{\pi^2}{k^3}\int_{2x_0}^{\infty} uJ_{p+1/2}(u)H_{p+1/2}^{(1)}(u)g(u/k)du, \label{eq:sp}
\end{eqnarray}
where ($a_n$, $b_n$) are the scattering coefficients of a spherical monomer obtained by the Lorenz--Mie theory \citep{BH83}, $\nstop$ is the truncation order of the scattering coefficients, $P_p$ is the Legendre polynomial function, and $P_n^m$ is the associated Legendre function, $J_{p+1/2}$ and $H_{p+1/2}^{(1)}$ are the Bessel and Hankel functions, respectively.
We have changed the lower bound of the integration in Equation (\ref{eq:sp}) from zero to $2x_0$ to make the formulation consistent with that in Section \ref{sec:ana}.

\subsection{Solution to mean-field equations in short-wavelength limit}
In general, the mean-field scattering coefficients ($\bar{d}_{1,n}^{(1)}, \bar{d}_{1,n}^{(2)}$) cannot be obtained analytically. 
However, there is an exceptional case that yields an analytic solution.  
First of all, we impose a wavelength much shorter than both monomer and aggregate radii so that $x_0=kR_0\gg1$ and $x_g=kR_g\gg1$.
Also, since we aim to evaluate geometric cross-sections, it is useful to assume perfectly absorbing monomers. With these assumptions, we derive an analytic solution to the extinction cross-section.

For sufficiently large dust aggregates ($x_g\gg1$), the asymptotic forms of the Bessel and Hankel functions yield $J_{p+1/2}(u)H_{p+1/2}^{(1)}(u)\sim(\pi{u})^{-1}$. Thus, Equation (\ref{eq:sp}) becomes
\begin{eqnarray}
s_p(kR_g)&\sim&\frac{\pi}{k^3}\int_{2x_0}^{\infty}g(u/k)du,\label{eq:spapp}\\
&\simeq&\frac{\tilsm}{x_0^2}, \label{eq:spsig}
\end{eqnarray}
where we used Equation (\ref{eq:tsmint}).
Because $s_p$ no longer depends on index $p$, $\bar{A}_{1,n}^{1,\nu}$ and $\bar{B}_{1,n}^{1,\nu}$ are reduced to a rather simple form \citep{B97}\footnote{Equation (21) in \citet{B97} contains errors and should be corrected in this way.}:
\begin{equation}
\bar{A}_{1,n}^{1,\nu}=\bar{B}_{1,n}^{1,\nu}=(2\nu+1)s_p\equiv f_\nu s_p. \label{eq:AB}
\end{equation}

For $x_0\gg1$, the Lorenz-Mie scattering coefficients are decomposed into two components:
\begin{eqnarray}
a_n&=&\frac{1}{2}\left(1+\sum_{l=0}^\infty a_n^{(l)}\right),\\
b_n&=&\frac{1}{2}\left(1+\sum_{l=0}^\infty b_n^{(l)}\right),
\end{eqnarray}
where the first term represents Fraunhofer diffraction, and the second term represents the sum with respect to Fresnel reflection ($l=0$: external reflection, $l=1$: transmitted light, $l\ge2$: internally reflected light) \citep[e.g.,][]{VDH57,T20}. From the assumption of perfectly absorbing monomers ($a_n^{(l)}=b_n^{(l)}=0$), the scattering coefficients become
\begin{equation}
a_n=b_n=\frac{1}{2}. \label{eq:diff}
\end{equation}
To highlight the meaning of Equation (\ref{eq:diff}), we consider the extinction cross-section of an isolated monomer:
\begin{eqnarray}
C_{\mathrm{ext}}^{\mathrm{mono}}&=&\frac{2\pi}{k^2}\sum_{n=1}^{\nstop}(2n+1)\mathrm{Re}(a_{n}+b_{n}),\\
&\simeq&\frac{2\pi}{k^2}\frac{2\nstop^2}{2},\\
&=&2\pi R_0^2, \label{eq:ext0}
\end{eqnarray}
where we used $\nstop\sim x_0$, as guaranteed by the localization principle \citep{VDH57}. Likewise, it is straightforward to show that the absorption and scattering cross-sections are $\pi R_0^2$, respectively \citep[e.g.,][]{BH83}. As a result, the extinction cross-section of an isolated monomer is twice its geometric cross-section, and the half comes from Fraunhofer diffraction and the other half from absorption.

Equation (\ref{eq:diff}) makes the mean-field equations exactly symmetric for the two modes, leading to a solution of $\bar{d}_{1,n}^{(1)}=\bar{d}_{1,n}^{(2)}$. Therefore, 
\begin{equation}
\bar{d}_{1,n}^{(1)}=\frac{1}{2}\left(1-2L\sum_{\nu=1}^{\nstop} f_\nu\bar{d}_{1,\nu}^{(1)}\right), \label{eq:dsolo}
\end{equation}
where $L=(N-1)s_p$. 
Equation (\ref{eq:dsolo}) can be solved analytically, and we obtain
\begin{equation}
\bar{d}_{1,n}^{(1)}=\bar{d}_{1,n}^{(2)}=\frac{1}{2}\frac{1}{1+L\sum_{\nu=1}^{\nstop}f_\nu}, \label{eq:sol}
\end{equation}
Using $\sum_{\nu=1}^{\nstop} f_\nu \simeq x_0^2$ and Equations (\ref{eq:spsig}) and (\ref{eq:sol}), Equation (\ref{eq:Cextmft}) becomes
\begin{equation}
C_\mathrm{ext}=\frac{2\pi NR_0^2}{1+(N-1)\tilsm}. \label{eq:Cextlim}
\end{equation}
Equation (\ref{eq:Cextlim}) is independent of the wavenumber $k$, indicating that the expression is in the geometrical optics limit.
Since we assumed a perfectly absorbing aggregate, we can anticipate $C_\mathrm{ext}=2G$ as a consequence of diffraction and absorption. Therefore, the geometric cross-section is given by
\begin{equation}
\frac{G}{N\pi R_0^2}=\frac{1}{1+(N-1)\tilsm}.\label{eq:mftfin}
\end{equation}
Equation (\ref{eq:mftfin}) is identical to Equation (\ref{eq:sigmean}).

Consequently, our analytic expression for geometric cross-sections is identical to the half of the extinction cross-section in the mean-field theory in the geometrical optics limit.

\section{Discussion} \label{sec:disc}

\subsection{Application to geometric cross-sections of QBCCA clusters}
\citet{O09} proposed another type of aggregation called quasi-BCCA (QBCCA). Here, we calculate geometric cross-sections of QBCCA clusters by extending our analytic expression to treat its inhomogeneous structure, and then, test its validity by comparing the measured cross-sections presented in \citet{O09}. 

The structure of a QBCCA cluster is specified by a parameter $\epsilon$ ($0<\epsilon\le1$), which determines a mass ratio of the two aggregates in each collision. QBCCA of $\epsilon=1$ corresponds to BCCA. 
For a given value of $\epsilon$, QBCCA is identical to BPCA when $N\le1.5/\epsilon$.
On the other hand, when $N>1.5/\epsilon$, a QBCCA cluster exhibits a fractal dimension of $D(\epsilon)$ (QBCCA regime). The fractal dimension increases with decreasing $\epsilon$, i.e., $D(\epsilon) \simeq1.92$ and $2.05$ for $\epsilon=0.325$ and $0.05$, respectively.
Thus, a QBCCA cluster has inhomogeneous structure: the small scale structure is relatively dense, whereas the large scale structure is relatively fluffy.

Our analytic expression (Equation \ref{eq:final}) is based on a homogeneous fractal cluster, and hence, it is not directly applicable to QBCCA clusters. To mimic the inhomogeneous structure, we adopt the following expressions
\begin{equation}
\frac{G}{N\pi R_0^2} =
\begin{dcases}
12.5N^{-0.315}\exp\left(-2.53/N^{0.0920}\right) & (N<N_\mathrm{th}),\\
\frac{A_1}{1+(N-1)\tilsm^{(\mathrm{BPCA})}} & (N_\mathrm{th}\le N\le1.5/\epsilon),\\
\frac{A_2}{1+(N-1)\tilsm} & (1.5/\epsilon < N),
\end{dcases} \label{eq:qbcca}
\end{equation}
where $A_1$ and $A_2$ are numerical factors to connect each regime continuously, and $\tilsm^{(\mathrm{BPCA})}$ is the overlapping efficiency of BPCA. 
The first and second expressions are identical to Equation (\ref{eq:final}) for BPCA. The third expression represents the cross-section at the QBCCA regime. If $1.5/\epsilon\le N_\mathrm{th}$, we directly connect the first and third expressions at $N=N_\mathrm{th}$.

We consider four QBCCA models: $\epsilon=0.325,~0.1,~0.05,~0.01$. 
We first determine the fractal dimension $D(\epsilon)$ by fitting the radius of gyration of each model using $R_g\propto N^{1/D(\epsilon)}$ at the QBCCA regime, and then, the fractal prefactor by using Equation (\ref{eq:k0app}) for a given value of $D(\epsilon)$. 
We do not use a value of $k_0$ directly measured from Equation (\ref{eq:Rg}) because this value is affected by the small-scale dense structure and is not a good indicator of the overlapping efficiency of the QBCCA regime.

\begin{figure}
\begin{center}
\includegraphics[width=\linewidth]{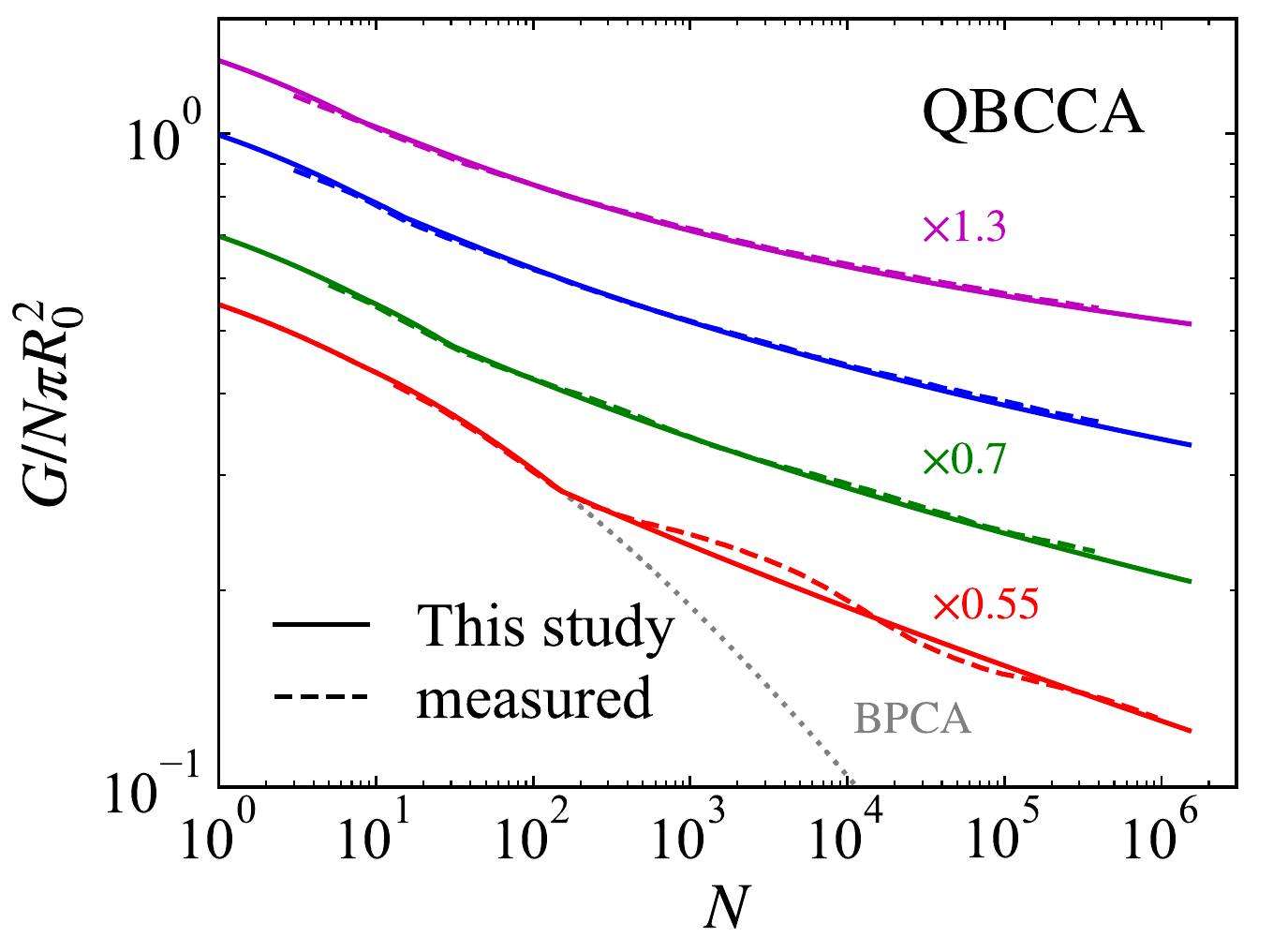}
\caption{Geometric cross-sections of aggregates formed via QBCCA. Solid and dashed lines represent the results obtained by Equation (\ref{eq:qbcca}) and numerically measured by \citet{O09}, respectively. From top to bottom, $\epsilon=0.325$, 0.1, 0.05, and 0.01, respectively, where $\epsilon$ represents a mass ratio of the two aggregates in each collision. For $\epsilon=0.01$, the cross-sections of BPCA clusters are also shown in the grey dotted line, which coincides with the red solid line at $N\le 1.5/\epsilon$. For clarity, each line is vertically shifted by a factor indicated in the figure.}
\label{fig:qbcca}
\end{center}
\end{figure}

Figure \ref{fig:qbcca} compares geometric cross-sections of the QBCCA clusters obtained by Equation (\ref{eq:qbcca}) and numerically measured by \citet{O09}. 
We find excellent agreement between the two results.
For $\epsilon\ge0.05$, the relative error is less than $2.1$ per cent.
For $\epsilon=0.01$, the measured cross-sections show oscillatory behaviour; nevertheless, our expression reproduces these cross-sections within the error of only 5.7 per cent.

The excellent agreement in Figure \ref{fig:qbcca} demonstrates that our formulation is valid for a fractal dimension between 1.9 (BCCA) and 3.0 (BPCA). Also, our analytic expression can be successfully extended to the case of an inhomogeneous fractal cluster by employing a simple prescription, as we employed Equation (\ref{eq:qbcca}) for QBCCA clusters.

\subsection{Comparison with commonly used estimates of cross-sections}

Model calculations of fractal grain growth often adopt various estimates of geometric cross-sections of fractal aggregates \citep[e.g.,][]{Cabane93, Wolf10, O12, Krijt15, Ohno20a, Ohno20b}. Here, we discuss how our analytically derived expression (Equation \ref{eq:final}) differs from various empirical formulas in the literature. 

\subsubsection{Characteristic cross-sections} \label{sec:gc}

Characteristic cross-sections (Equation \ref{eq:pirc2}), or the almost similar expression $\pi R_g^2$, have been frequently used in the literature to estimate geometric cross-sections \citep[e.g.,][]{Cabane93, Wolf10, Ohno20a}. As mentioned in Section \ref{sec:sl}, Equation (\ref{eq:pirc2}) becomes unphysical when $D<2$. Thus, we adopt a simple prescription:
\begin{equation}
\frac{G}{N\pi R_0^2} = \mathrm{min}\left[\frac{1}{N}\left(\frac{R_c}{R_0}\right)^2,1\right]. \label{eq:pirc2b}
\end{equation}

\begin{figure*}
\begin{center}
\includegraphics[width=0.49\linewidth]{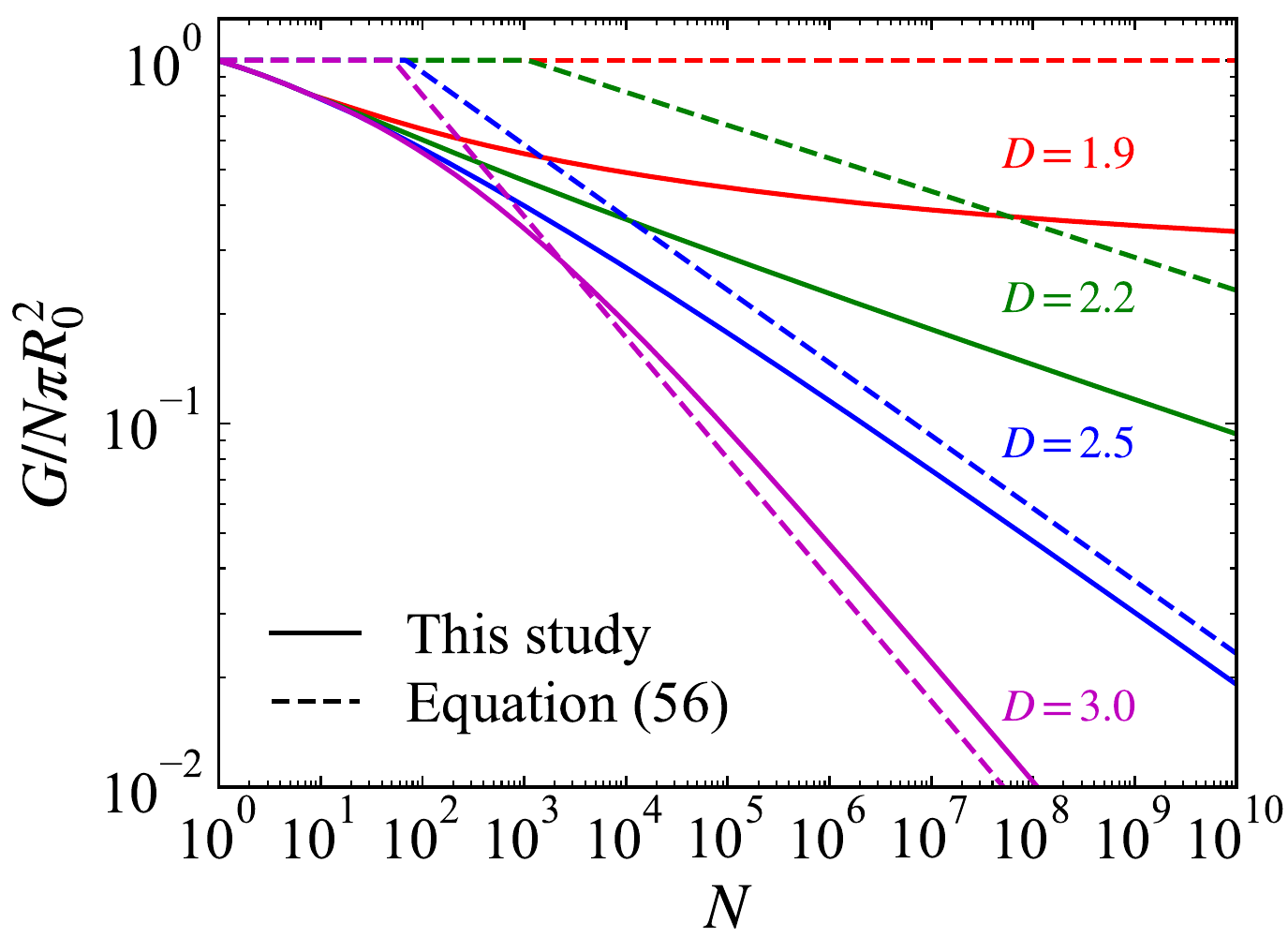}
\includegraphics[width=0.49\linewidth]{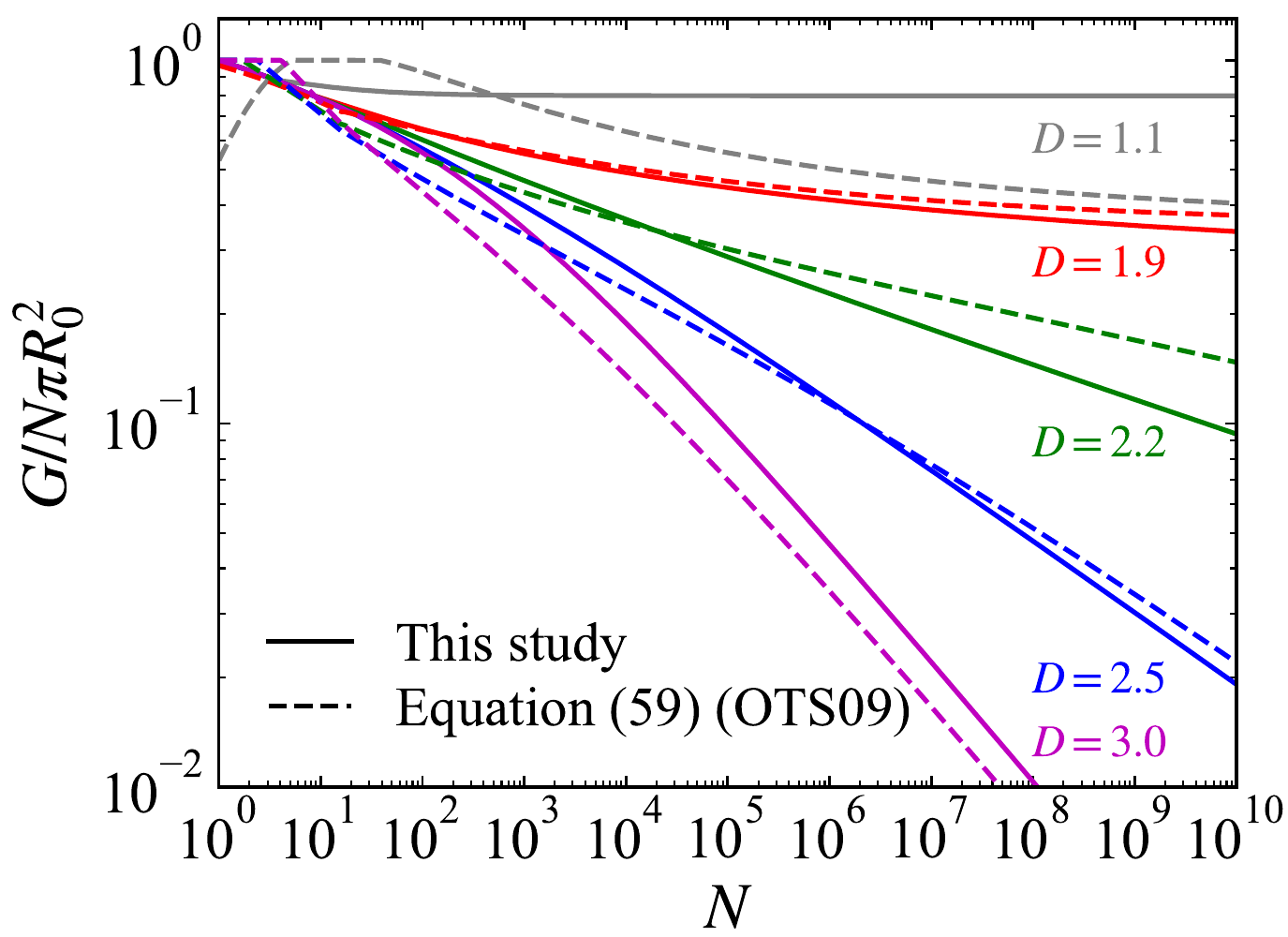}
\caption{Comparison of geometric cross-sections between our analytically derived expression (Equation \ref{eq:final}, solid lines) and commonly used formulas (dashed lines). The left and right panels show the results obtained by the characteristic cross-section (Equation \ref{eq:pirc2b}) and the empirical formula in \citet{O09} (Equation \ref{eq:O09}), respectively. Red, green, blue, and purple lines represent the results for $D=1.9$, $2.2$, $2.5$, and $3.0$, respectively. In the right panel, the result for $D=1.1$ is also shown. The fractal prefactor is determined by Equation (\ref{eq:k0app}). }
\label{fig:pre}
\end{center}
\end{figure*}

Figure \ref{fig:pre} compares our analytic expression (Equation \ref{eq:final}) with Equation (\ref{eq:pirc2b}), and Figure \ref{fig:errormap} shows the relative error between them.
Equation (\ref{eq:pirc2b}) shows moderate agreement with our analytic expression when $D=3.0$. 
To compare the two results, we introduce the area-equivalent radius: $R_a=\sqrt{G/\pi}$.
At $N=10^3$, the area-equivalent radius differs from the characteristic radius by only 4 per cent, which is quantitatively consistent with measurements reported in \citet{O09}. However, the difference increases with increasing $N$.
For sufficiently large $N$, using Equation (\ref{eq:tilsig_Q1}), we have
\begin{equation}
\frac{R_a}{R_c}\simeq 2^{\frac{1+D}{D}}\sqrt{A\frac{3/5}{\Gamma\left(\frac{D-2}{D}\right)}}\simeq1.14~ (D=3.0), \label{eq:radf3}
\end{equation}
where we substituted $A\simeq0.917$ for BPCA. Therefore, the area-equivalent radius of BPCA is approximately 14 per cent larger than the characteristic radius at large $N$, and then a relative error of $G$ becomes $\sim23$ per cent. 

The relative error of Equation (\ref{eq:pirc2b}) increases with decreasing $D$ for $2< D\le3$. In this range of fractal dimensions, both our expression and Equation (\ref{eq:pirc2b}) fulfill the scaling law (Equation \ref{eq:scaling}); however, Equation (\ref{eq:pirc2b}) significantly overestimates the cross-sections as $D$ approaches 2.0.
For $D=2.0$, the relative error of the characteristic cross-section exceeds $100$ per cent at $N\sim10^3$, and the error monotonically increases with rising $N$. This is because Equation (\ref{eq:pirc2b}) has a constant value, whereas our expression has logarithmic dependence (Section \ref{sec:asym2}). For $D<2$, the relative error decreases with decreasing $D$ at large $N$. For $D=1.0$, the relative error is approximately 18 per cent at large $N$.

Therefore, the characteristic cross-section contains the relative error of $\sim20$ per cent when $D=1.0$ and $D=3.0$, whereas the error is particularly pronounced at $D\sim2$ at large $N$. 

\subsubsection{\citet{P09}}

\citet{P09} realized that fluffy aggregates satisfy the following empirical relation:
\begin{equation}
N\left(\frac{R_0}{R_a}\right)^3=1.21\left(\frac{R_\mathrm{out}}{R_a}\right)^{-0.3}N^{-0.33}, \label{eq:PD09}
\end{equation}
where $R_\mathrm{out}$ is the outermost radii of aggregates. This relation is valid when $R_\mathrm{out}/R_a>1.2$, indicating that it is not applicable to compact aggregates, i.e., higher values of $D$. 

We compared our analytic expression with Equation (\ref{eq:PD09}), where we approximate $R_\mathrm{out}$ as $R_c$. We found that Equation (\ref{eq:PD09}) tends to underestimate (overestimate) the cross-section when $D$ is smaller (larger) than $D\simeq2.3$. This is consistent with the findings presented in previous studies \citep{O09, S12}. \citet{O09} pointed out that Equation (\ref{eq:PD09}) underestimates the cross-sections of QBCCA clusters, particularly when $D\lesssim2.05$ ($\epsilon=0.05$).
\citet{S12} conducted another test by measuring the cross-sections of compressed aggregates ($D\simeq2.5$) and concluded that Equation (\ref{eq:PD09}) causes overestimation. Thus, our findings are consistent with the previous studies.

\subsubsection{\citet{O09}}
\begin{figure*}
\begin{center}
\includegraphics[width=0.49\linewidth]{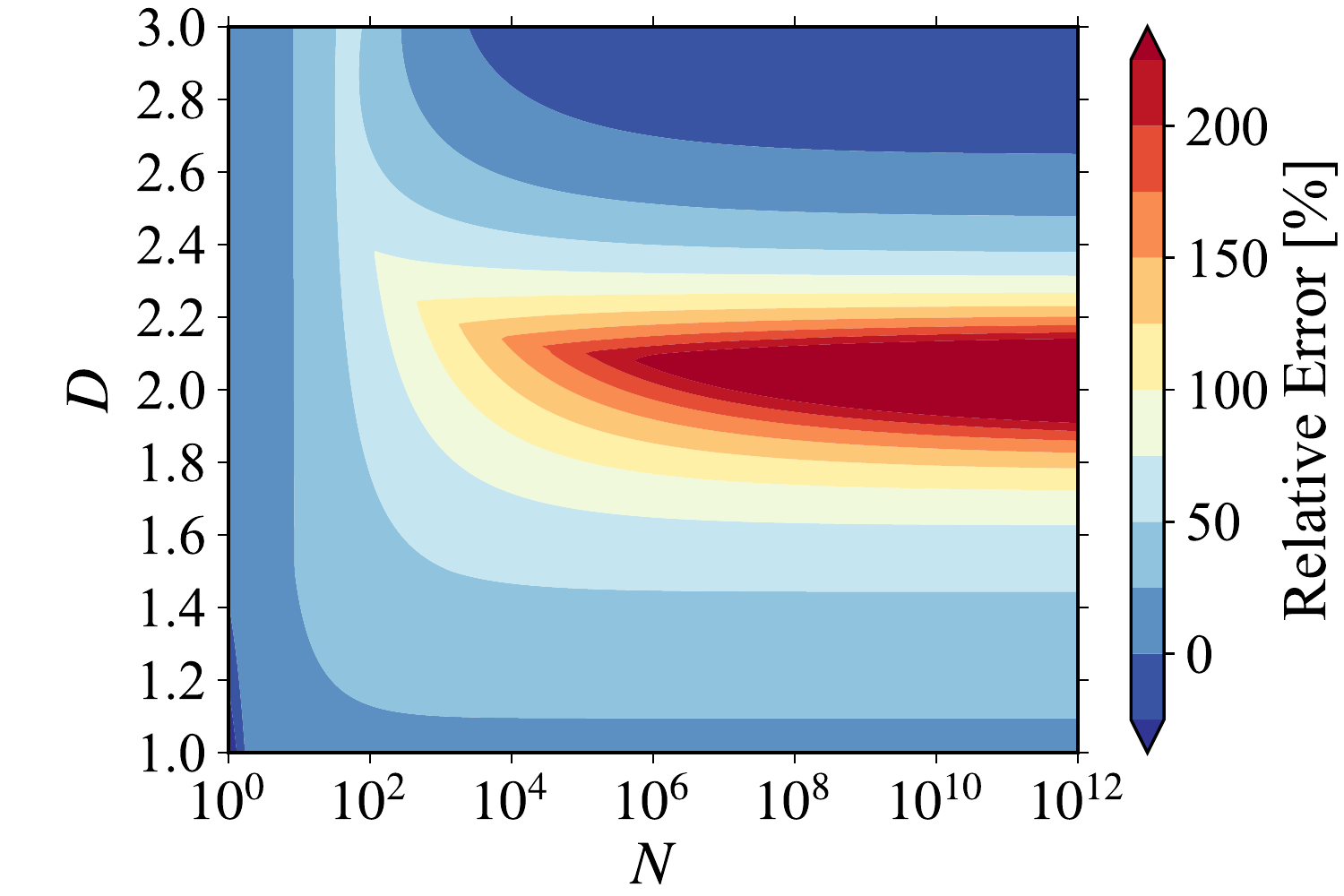}
\includegraphics[width=0.49\linewidth]{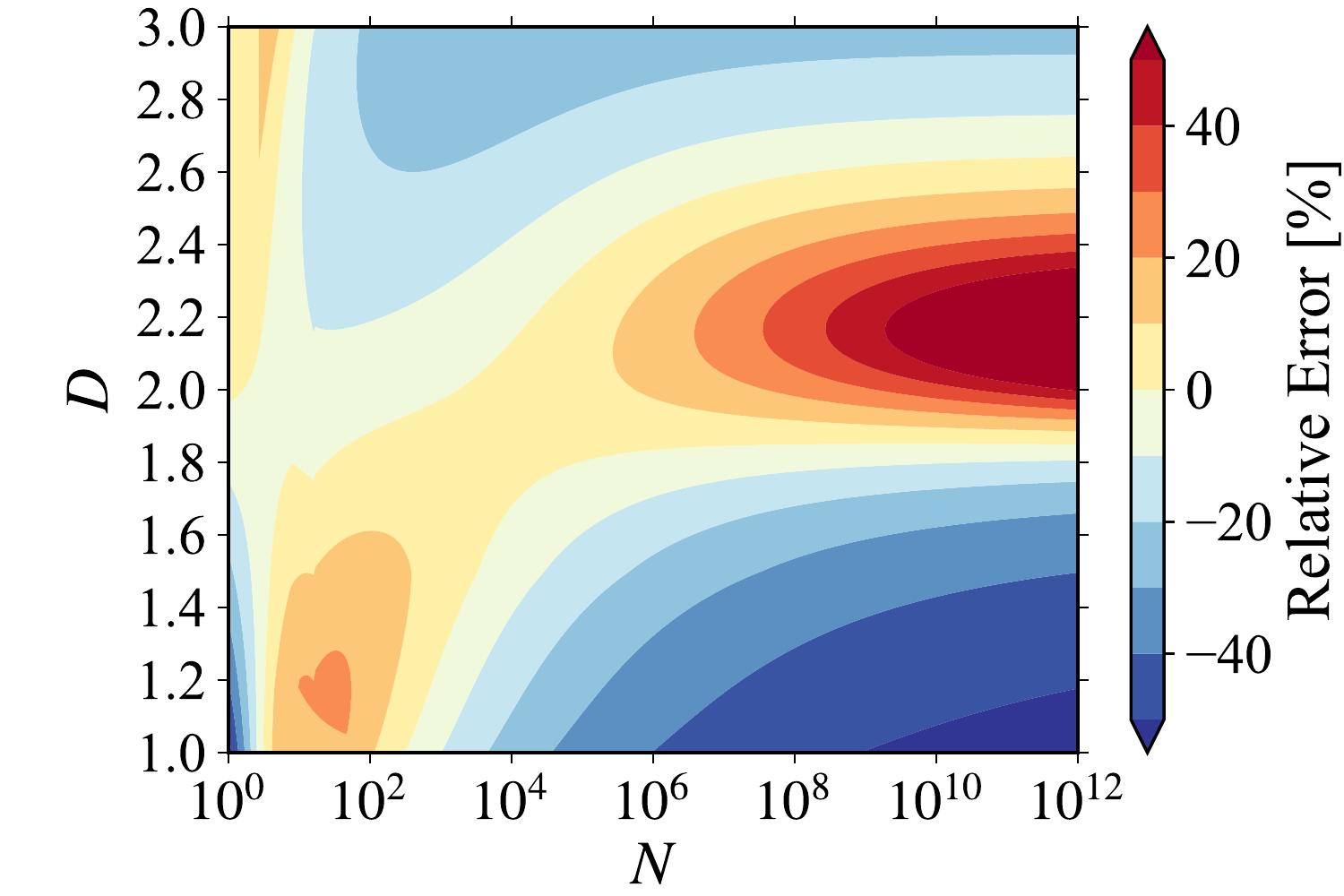}
\caption{Relative errors of the characteristic cross-section (Equation \ref{eq:pirc2b}) (left) and the empirical formula in \citet{O09} (Equation \ref{eq:O09}) (right). Positive and negative values correspond to cases where these commonly used formulas over- and under-estimate the cross-section in comparison with our analytic expression, respectively.}
\label{fig:errormap}
\end{center}
\end{figure*}

\citet{O09} proposed another empirical formula, which provides a better fit to measured cross-sections of QBCCA clusters than Equation (\ref{eq:PD09}). This formula has been widely used in model calculations of fractal grain growth in protoplanetary discs and planetary atmospheres \citep[e.g.,][]{O12,Krijt16, Ohno20b}.
The empirical formula is given by
\begin{equation}
G(N,R_c)=\left(\frac{1}{G_\mathrm{BCCA}}+\frac{1}{\pi R_c^2}-\frac{1}{\pi R_{c,\mathrm{BCCA}}^2}\right)^{-1}, \label{eq:O09}
\end{equation}
where $G_\mathrm{BCCA}$ is the cross-section of a BCCA cluster, obtained from the fitting formulas in \citet{Minato06} (Equations \ref{eq:minatosmall} and \ref{eq:minato}), and $R_{c,\mathrm{BCCA}}$ is the characteristic radius of the BCCA cluster. At sufficiently large $N$, Equation (\ref{eq:O09}) yields $G\simeq G_\mathrm{BCCA}$ in the BCCA limit and $G\simeq\pi R_c^2$ in the BPCA limit. 
Because Equation (\ref{eq:O09}) sometimes yields $G>N\pi R_0^2$ at small $N$, we set $G=N\pi R_0^2$ in such cases.

We compare our analytic expression with the empirical formula (Equation \ref{eq:O09}) in Figure \ref{fig:pre}. 
We also show the relative error between them in Figure \ref{fig:errormap}. 
As an overall tendency, the empirical formula agrees with our analytic expression within the error of 40 per cent for small aggregates ($N<10^6$) of an arbitrary fractal dimension. In contrast, the error tends to increase for larger aggregates ($N>10^6$) particularly at $D\sim1$ and $\sim2.2$. The two methods are in good agreement for $D=1.9$, since both of them can reproduce the cross-sections of BCCA clusters. 

Main differences between the empirical formula (Equation \ref{eq:O09}) and our analytic expression are as follows.
\begin{itemize}
\item $D=1.1$:
The empirical formula underestimates a cross-section when $N\lesssim3$ and $N\gtrsim500$. At large $N$, the relative error gradually increases with rising $N$ and exceeds 20 per cent for $N\gtrsim10^4$, as the term $G_\mathrm{BCCA}$ tends to govern the reciprocal sum.

\item
$D=2.2$: The empirical formula underestimates a cross-section by at most about 11 per cent when $N\lesssim10^4$. A similar underestimation has also been reported in \citet{O09} for QBCCA clusters. When $N\gtrsim10^4$, the empirical formula overestimates a cross-section because it does not follow the scaling law (Equation \ref{eq:scaling}). 

\item $D=2.5$:  
The empirical formula underestimates a cross-section by at most about $17$ per cent when $N\sim10^2-10^3$. A similar difference has been reported in \citet{S12} for compressed aggregates ($D\simeq2.5$) at $N\sim10^2-10^3$. 

\item $D=3.0$:  
Since the empirical formula approaches $\pi R_c^2$ at large $N$, the relative error is almost the same as that of the characteristic cross-section (Section \ref{sec:gc}). However, the empirical formula is slightly less accurate than $\pi R_c^2$ at small aggregates ($N\sim10^3$), i.e., the relative error is approximately 28 per cent at $N\sim5\times10^3$. This is because the reciprocal sum in the formula does not accurately yield $G\sim \pi R_c^2$ for such a small $N$, which causes larger errors than $\pi R_c^2$.
\end{itemize}

In summary, Equation (\ref{eq:O09}) tends to produce better results among other empirical formulas; however, it leads to significant errors for very large aggregates ($N\gtrsim10^6$). Also, Equation (\ref{eq:O09}) fails to reproduce the scaling law at $D\sim2.2$. 

\section{Summary} \label{sec:con}
We derived an analytic expression for geometric cross-sections of fractal dust aggregates by applying a statistical distribution model of monomers.
Our main results are as follows.
\begin{itemize}
\item Our analytic expression (Equation \ref{eq:final}) successfully reproduces the cross-sections of three types of fractal aggregates (the linear chain, BCCA, and BPCA clusters) with a relative error below 3 per cent. Furthermore, our expression naturally reproduces the scaling law (Equation \ref{eq:scaling}) at sufficiently large $N$.

\item The analytic expression is shown to be identical to an expression predicted by the mean-field light scattering theory in the short-wavelength limit. Therefore, our formulation for geometric cross-sections is compatible with the mean-field theory.

\item We extended our analytic expression to calculate the cross-sections of QBCCA clusters, which exhibit inhomogeneous structure. The cross-sections obtained by the extended expression show excellent agreement with the measured cross-sections, where the error is only less than 5.7 per cent. This agreement suggests that our formulation is valid even for a case of a fractal dimension between 1.9 (BCCA) and 3.0 (BPCA). 
 
\item While the empirical formula in \citet{O09} is the best among various fitting formulas in the literature, it leads to significant errors for very large aggregates ($N\gtrsim10^6$) and fails to reproduce the scaling law (Equation \ref{eq:scaling}). 

\end{itemize}

Our analytic expression comes at a low computational cost and yields better accuracy than the empirical formulas proposed in previous studies. Therefore, it is useful in model calculations of fractal grain growth in protoplanetary discs and planetary atmospheres. 

\section*{Acknowledgements}
We thank Robert Botet and Carsten Dominik for fruitful discussions. 
We also thank Satoshi Okuzumi and Hidekazu Tanaka for providing the cross-section data of QBCCA clusters. R.T. acknowledges JSPS overseas research fellowship. This work was supported by JSPS KAKENHI Grant Numbers JP19H05068.

\section*{Data Availability}
The data underlying this article are available in Zenodo at \url{https://doi.org/10.5281/zenodo.4542794}.




\bibliographystyle{mn2e}



\appendix
\section{Geometric cross-sections of linear chain clusters ($D=1$)} \label{sec:chain}
Assuming a linear chain cluster consisting of $N$ monomers and denoting the angle between a light ray and  the linear chain cluster by $\Theta$,
we calculate the shadow area cast by this cluster onto the plane perpendicular to the light ray direction. To calculate the shadow area, we consider a rectangle circumscribing the shadow in the projection plane. Each length of the rectangle will be $2R_0$ and $2R_0[1+(N-1)\sin\Theta]$, and hence, the area of the rectangle is given by $4R_0^2[1+(N-1)\sin\Theta]$. The shadow area is obtained by subtracting the area of the marginal regions between the rectangle and the shadow from that of the rectangle.
Thus, the shadow area $g(\Theta)$ is expressed as
\begin{eqnarray}
g(\Theta) &=& 4R_0^2[1+(N-1)\sin\Theta] - \nonumber\\
&&4R_0^2\left[\left(1-\frac{\pi}{4}\right)+(N-1)\left(\sin\Theta-\frac{\pi}{4}+\frac{\sigma_{ij}}{4R_0^2}\right)\right],\nonumber\\
&=& \pi R_0^2 \left\{1+(N-1)\left[1-\frac{2}{\pi}\arcsin(\cos\Theta)+\frac{2}{\pi}\sin\Theta\cos\Theta\right]\right\},\nonumber\\
\end{eqnarray}
where we employed Equation (\ref{eq:sigij}) with $\rho=1$.
Evidently, $g(0)=\pi R_0^2$ and $g(\pi/2)=N\pi R_0^2$.
The average shadow area of randomly orientated linear chain clusters is given by
\begin{equation}
\frac{G}{N\pi R_0^2}=\int_0^{\pi/2} d\Theta \frac{g(\Theta)}{N\pi R_0^2} \sin\Theta.
\end{equation}
Therefore, we obtain
\begin{equation}
\frac{G}{N\pi R_0^2}=\frac{1}{N}+\left(1-\frac{1}{N}\right)\frac{8}{3\pi}.
\end{equation}

\section{Effect of cut-off functions of correlation function} \label{sec:cutoff}
\begin{figure*}
\begin{center}
\includegraphics[width=0.49\linewidth]{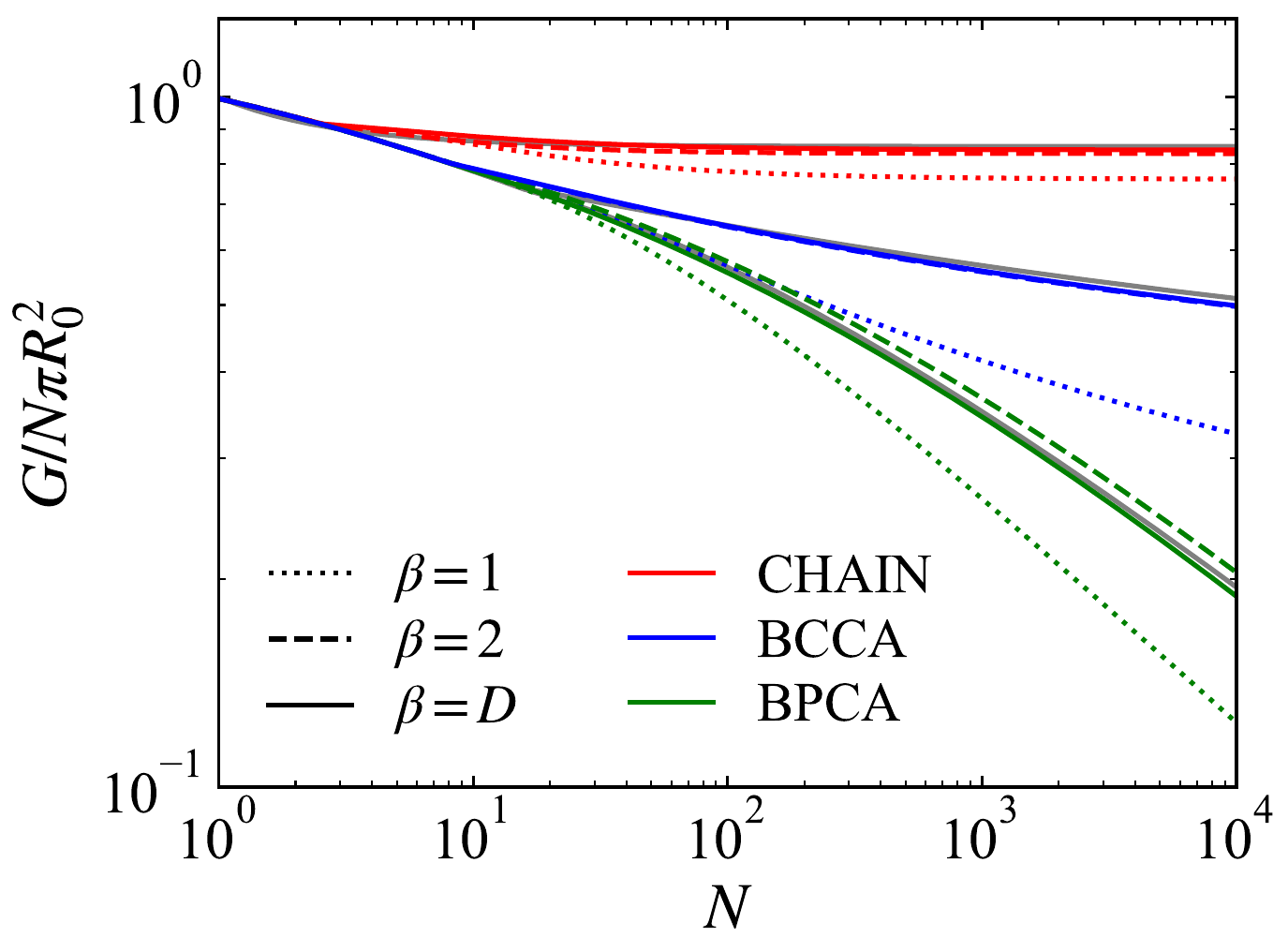}
\includegraphics[width=0.49\linewidth]{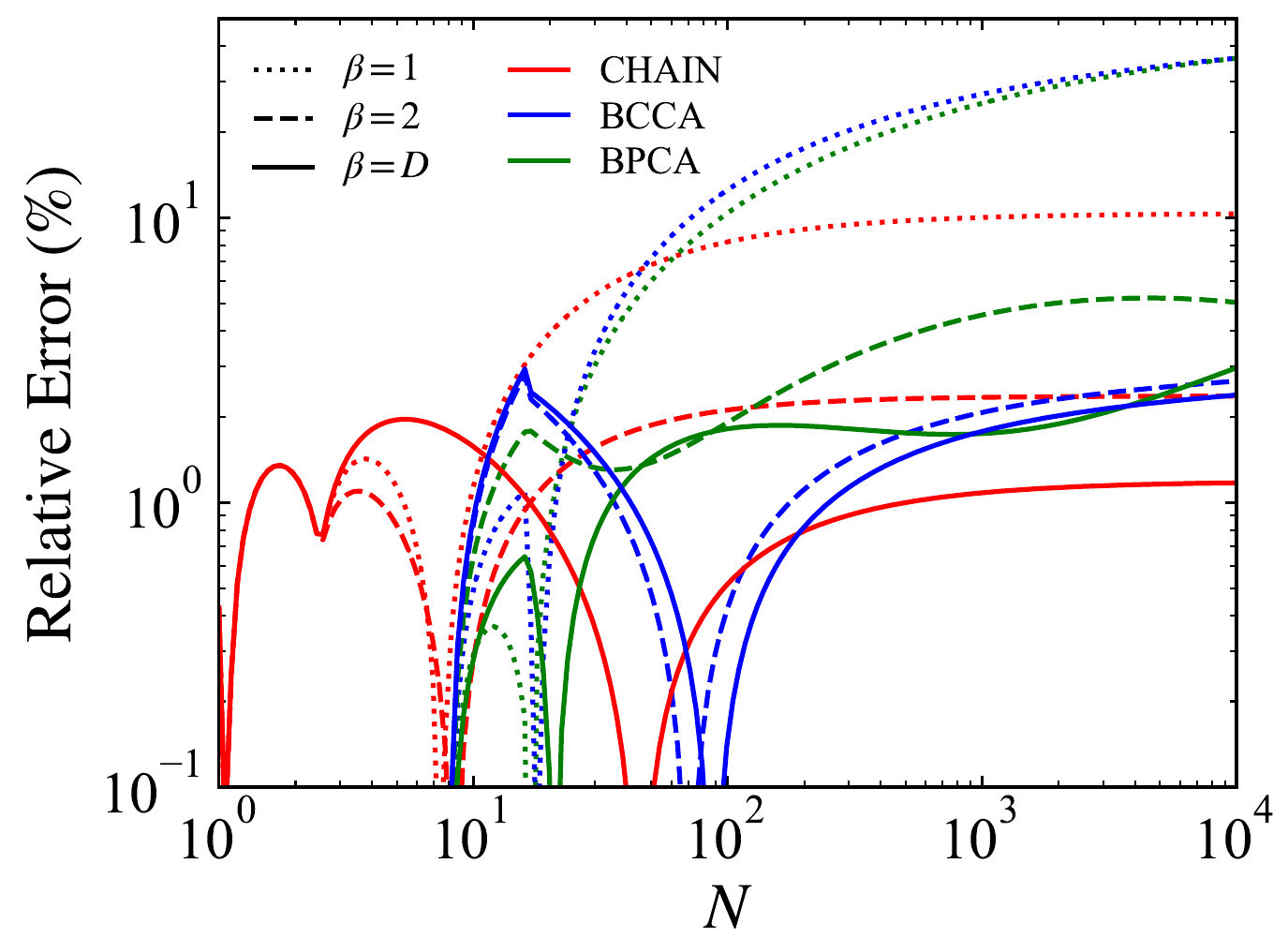}
\caption{Comparison between three cut-off models: geometric cross-sections of fractal aggregates (left) and relative errors of our analytic expression (right). Dotted, dashed, and solid lines represent the results for $\beta$=1, 2, and $D$, respectively. The red, blue, and green lines represent the results for the linear chain, BCCA, and BPCA clusters, respectively. The grey solid lines in the left panel represent numerically measured or exact cross-sections of BCCA, BPCA, and linear chain clusters (Equations \ref{eq:minatosmall}, \ref{eq:minato}, \ref{eq:chainexact}).}
\label{fig:cutoff}
\end{center}
\end{figure*}

In Section \ref{sec:derive}, we derived geometric cross-sections of aggregates by assuming a form of the two-point correlation function. Because an aggregate has a finite radius, the two-point correlation function must have a cut-off function at a length scale comparable to the aggregate radius. The cut-off function has been commonly assumed to have the following form $f_c(r/R_g)\propto \exp[-(r/R_g)^\beta]$, where $\beta$ represents the cut-off power. In our fiducial model, we adopted $\beta=D$ (see Section \ref{sec:derive}) in accordance with previous studies \citep{B95, T18}. Here, we investigate how different values of $\beta$ affect calculations of geometric cross-sections.

\citet{Berry86} proposed the exponential cut-off model ($\beta=1$), given by
\begin{equation}
f_c\left(\frac{r}{R_g}\right)=\frac{\left[D(D+1)/2\right]^{D/2}}{\Gamma(D)}\exp\left\{-\left[\frac{D(D+1)}{2}\right]^\frac{1}{2}\left(\frac{r}{R_g}\right)\right\}, \label{eq:comodel_ex}
\end{equation}
A similar analysis given in Section \ref{sec:derive} yields the following analytic expressions for the overlapping efficiency:
\begin{equation}
\tilsm=\frac{\eta_1^2}{16\Gamma(D)}
\begin{dcases}
\Gamma(D-2)Q(D-2,\eta_1) & (2<D\le3),\\
\int_{\eta_1}^\infty dx x^{D-3}e^{-x} & (D\le2),
\end{dcases}
\end{equation}
where $\eta_1=2(R_0/R_g)[D(D+1)/2]^{1/2}$.
Another cut-off model is the Gaussian cut-off model ($\beta=2$) \citep[e.g.,][]{Sorensen92, T16}, which has the form of 
\begin{equation}
f_c\left(\frac{r}{R_g}\right)=\frac{2^{1-D}D^{D/2}}{\Gamma(D/2)}\exp{\left[-\frac{D}{4}\left(\frac{r}{R_g}\right)^2\right]}.\label{eq:comodel_gc}
\end{equation}
In this case, the overlapping efficiency can be expressed as
\begin{equation}
\tilsm=\frac{\eta_2}{16\Gamma(D/2)}
\begin{dcases}
\Gamma\left(\frac{D-2}{2}\right)Q\left(\frac{D-2}{2},\eta_2\right) & (2<D\le3),\\
\int_{\eta_2}^\infty dx x^{\frac{D}{2}-2}e^{-x} & (D\le2),
\end{dcases}
\end{equation}
where $\eta_2=D(R_0/R_g)^2$.

Figure \ref{fig:cutoff} compares geometric cross-sections obtained by three different cut-off models. For our fiducial model ($\beta=D$), the geometric cross-sections agree with the numerically measured and exact cross-sections (Equations \ref{eq:minatosmall}, \ref{eq:minato}, \ref{eq:chainexact}) within the error of $3$ per cent. For the Gaussian cut-off model ($\beta=2$), the relative error is below $\sim5$ per cent. However, for the exponential cut-off model ($\beta=1$), the relative error is above 30 per cent. Therefore, we do not recommend to use $\beta=1$. These findings are consistent with the conclusion derived in \citet{T18}, where the authors compared extinction cross-sections of BCCA and BPCA clusters obtained by a rigorous light scattering simulation with those obtained by the mean-field light scattering theory.

\bsp	
\label{lastpage}
\end{document}